\documentclass{pasj00} \usepackage{times} 

\begin{document}
\SetRunningHead{B. Hatsukade \etal}{CO search in an sBzK at $z = 2.044$}

\Received{2008/11/19}
\Accepted{2009/01/15}

\title{A Search for Molecular Gas toward a $BzK$-selected Star-forming Galaxy at $z = 2.044$}

\author{Bunyo \textsc{Hatsukade},\altaffilmark{1}
        Daisuke \textsc{Iono},\altaffilmark{2}
        Kentaro \textsc{Motohara},\altaffilmark{1}
        Kouichiro \textsc{Nakanishi},\altaffilmark{2,5}
        Masao \textsc{Hayashi},\altaffilmark{3}
        Kazuhiro \textsc{Shimasaku},\altaffilmark{3}
        Tohru \textsc{Nagao},\altaffilmark{4}
        Yoichi \textsc{Tamura},\altaffilmark{3,5}
        Matthew A \textsc{Malkan},\altaffilmark{6}
		Chun \textsc{Ly},\altaffilmark{6}
        and
        Kotaro \textsc{Kohno}\altaffilmark{1}
        }
\altaffiltext{1}{Institute of Astronomy, University of Tokyo, 2-21-1 Osawa, Mitaka, Tokyo 181-0015}
\email{hatsukade@ioa.s.u-tokyo.ac.jp}
\altaffiltext{2}{Nobeyama Radio Observatory, Minamimaki, Minamisaku, Nagano 384-1805}
\altaffiltext{3}{Department of Astronomy, Graduate School of Science, University of Tokyo, Tokyo 113-0033}
\altaffiltext{4}{Physics Department, Graduate School of Science and Engineering, Ehime University, 2-5 Bunkyo-cho, Matsuyama 790-8577}
\altaffiltext{5}{National Astronomical Observatory of Japan, 2-21-1 Osawa, Mitaka, Tokyo 181-8588}
\altaffiltext{6}{Department of Physics and Astronomy, University of California at Los Angeles, P. O. Box 951547, Los Angeles, CA, 90095-1547, USA}


\maketitle

\begin{abstract}
We present a search for CO~(3--2) emission in SDF-26821, a $BzK$-selected star-forming galaxy (sBzK) at $z = 2.044$, 
using the 45-m telescope of the Nobeyama Radio Observatory 
and the Nobeyama Millimeter Array. 
We do not detect significant emission and derive 2 $\sigma$ limits: 
the CO luminosity of $L'_{\rm{CO}} < 3.1 \times 10^{10}$ K km s$^{-1}$ pc$^{-2}$, 
the ratio of far-infrared luminosity to CO luminosity of $L_{\rm{FIR}}/L'_{\rm{CO}} > 57$ \LO\ (K km s$^{-1}$ pc$^{-2}$)$^{-1}$, and the molecular gas mass of $M_{\rm{H_2}} < 2.5 \times 10^{10}$ \MO, 
assuming a velocity width of 200 km s$^{-1}$ and a CO-to-H$_2$ conversion factor of $\alpha_{\rm{CO}}=0.8$ \MO\ (K km s$^{-1}$ pc$^{-2}$)$^{-1}$. 
The ratio of $L_{\rm{FIR}}/L'_{\rm{CO}}$, a measure of star formation efficiency (SFE), is comparable to or higher than the two $z \sim 1.5$ sBzKs detected in CO~(2--1) previously, suggesting that sBzKs can have a wide range of SFEs. 
Comparisons of far-infrared luminosity, gas mass, and stellar mass among the sBzKs suggest that SDF-26821 is at an earlier stage of forming stars with a similar SFE and/or more efficiently forming stars than the two $z \sim 1.5$ sBzKs. 
The higher SFEs and specific star formation rates of the sBzKs compared to local spirals are indicative of the difference in star formation modes between these systems, suggesting that sBzKs are not just scaled-up versions of local spirals. 
\end{abstract}

\section{Introduction}
Molecular gas, fuel for star formation, is a fundamental component of galaxies. 
Observations of molecular gas, traced by CO emission lines, provide important information on the evolution of galaxies: e.g., star formation efficiency and the timescale and the formation phase of galaxies. 
In the formation history of galaxies in the universe, the era of $z \sim 1$--3 is thought to be a turning point, since the cosmic star formation rates peaked and the majority of stellar mass had formed by $z \sim 1$ (e.g., \cite{dick03}). 
Therefore it is important to reveal the gas environment of this era to probe the star formation activities in galaxies. 
However, the detection of CO line at this era has been made mainly in apparently bright galaxies such as radio galaxies, QSOs, and bright submillimeter galaxies (SMGs), or in lensed galaxies (see \cite{solo05} for a review). 
For a comprehensive understanding of the galaxy evolution, it is necessary to observe molecular gas in galaxies which dominate the galaxy population in the universe. 

The $BzK$ color selection technique using two colors, $B-z'$ and $z'-K$, provides a $K$-limited (i.e., mass-limited) sample of galaxies at $1.4 \lesssim z \lesssim 2.5$ \citep{dadd04}. 
This technique is less affected by dust extinction and effectively selects dusty starburst galaxies compared to samples of $z \sim 2$ UV-selected star-forming galaxies, BX and BM galaxies \citep{stei04}. 
$BzK$-selected galaxies (BzKs) are classified into two categories, star-forming galaxies (sBzKs) and passively-evolving galaxies (pBzKs), based on their colors. 
CO observations of BzKs and BXs have been performed by \citet{dadd08a} (hereafter D08) and \citet{tacc08}. 
D08 successfully detected CO~(2--1) emission from two sBzKs at $z \sim 1.5$, yielding molecular gas mass of $M_{\rm{H_2}}\sim 2 \times  10^{10}$ \MO. 
The large amount of molecular gas with relatively low stellar mass ($M_* = 5 \times 10^{10}$ and $8 \times 10^{10}$ \MO, adopting a \citet{chab03} initial mass function (IMF)) and high star formation rates (SFR$_{\rm{FIR}} = 130$ and 235 \MO\ yr$^{-1}$) imply that gas-rich galaxies in the process of formation prevail in the early universe, considering the fact that the volume density of sBzKs is a factor of 10 larger than that of SMGs (e.g., \cite{dadd04}; \cite{chap05}). 
On the other hand, \citet{tacc08} did not detect CO emission in a BzK and two BXs at $z \sim 2$, providing upper limits of molecular gas mass of 5--$7 \times 10^9$ \MO\ (2 $\sigma$). 
This is inconsistent with galaxies in the local universe because we expect a gas mass of a few $10^{10} \MO$ for a galaxy with SFR of $\sim 200$--300 \MO\ yr$^{-1}$ assuming a standard Schmidt law \citep{schm59, kenn98}. 

This may be caused by differences in physical properties of these systems such as CO-to-H$_2$ conversion factors, excitation conditions of CO, relation between gas surface density and SFR, and star formation modes. 
It is essential to increase the number of CO observations of these high-redshift star-forming galaxies to examine the physical properties. 

In this letter, we report a search for CO~(3--2) line emission in an sBzK detected in the Subaru Deep Field (SDF; \cite{kash04}). 
Our target, SDF-26821, is selected from deep $K$-band imaging \citep{haya07} and confirmed to be at $z=2.044\pm0.002$ by the Subaru/MOIRCS near-infrared spectroscopy (\cite{haya09}, hereafter H08).
The FIR luminosity is estimated to be $L_{\rm FIR}=1.7\times 10^{12} L_{\odot}$ from the 24 $\mu$m flux obtained with the {\it Spitzer}/MIPS observations (Ly et al. in prep.), using a template SED of local galaxies \citep{char01}, with about factor of a few  uncertainty \citep{papo07}. 
Thus, this galaxy is classified as an Ultra Luminous Infrared Galaxy (ULIRG).
The SFR is the largest among the spectroscopic sample: 
370, 730, and 300 \MO\ yr$^{-1}$ derived from rest-frame 1500 \AA\ corrected for dust extinction using the Calzetti extinction law, extinction-corrected H$\alpha$, and FIR luminosities, respectively (H08). 
Although there is a possible contribution from an active galactic nucleus (AGN), the contribution appears to be insignificant (H08). 
The stellar mass of $3.4 \times 10^{10} \MO$ is derived from the SED fitting (with about 30\% of fitting error) assuming a \citet{salp55} IMF with lower and upper mass cutoffs of 0.1 $M_{\odot}$ and 100 $M_{\odot}$ and a continuous burst.
The large SFR compared to the relatively small stellar mass suggests that SDF-26821 is in its forming phase and may contain a large amount of molecular gas. 

In \S 2 we describe our CO~(3--2) observations and data reduction. 
In \S 3 we present the results and physical quantities obtained. 
In \S 4 we compare CO luminosity, FIR luminosity, and stellar mass of SDF-26821, sBzKs, and other galaxy populations and discuss the difference in star formation. 
Where necessary, we adopt a cosmological parameter of $H_0=70$ km s$^{-1}$ Mpc$^{-1}$, $\Omega_{\rm{M}}=0.3$, and $\Omega_{\Lambda}=0.7$. 

\section{Observations and Data Reduction}
CO~(3--2) observations were initially conducted with the 45-m telescope of the Nobeyama Radio Observatory (NRO) on March 13--15, 2008. 
The observation frequency was set to the redshifted CO~(3--2) line of 113.599 GHz at the upper sideband of the T100 receiver, a waveguide-type dual-polarization sideband-separating SIS receiver \citep{naka08}. 
The S100 receiver was also used for gain calibrations of the T100 receiver. 
The half-power beam width (HPBW) of the telescope for two polarizations were $15''.5 \pm 0''.1$ and $17''.6 \pm 0''.3$, and the main-beam efficiency was 34\% and 35\%. 
We used the AOS-Wide spectrometer with 2048 channels and a bandwidth of 250 MHz. 
The pointing was checked every hour using SiO masers. 
The sky emission was subtracted by position switching. 
The intensity calibration was performed by the chopper wheel method, 
and the absolute intensity calibration was performed by observing IRC$+10216$. 
The system temperature was in the range of 300--500 K in SSB. 

We then conducted observations with the Nobeyama Millimeter Array (NMA) on May 19--27, 2008. 
We used the most compact array configuration, D-configuration (baseline length: 15--82~m), with five 10-m antennas. 
The Ultra Wide Band Correlator (UWBC; \cite{okum00}) was used with 128 channels and a bandwidth of 1024 MHz, covering $\sim 2700$ km s$^{-1}$.  
The bandwidth was wide enough even allowing for the uncertainty of observation frequency ($\sim 150$ MHz) due to the uncertainty of redshift. 
The observation frequency was set to the redshifted CO~(3--2) line of 113.599 GHz at the upper sideband, 200 MHz shifted downward from the center of the band to prevent spurious features. 
The field of view is $62''$ in diameter. 
The QSO 1308+326 was observed every 20 minutes for visibility calibrations, and 3C~273 were observed for bandpass calibrations. 
Data reduction was carried out using the UVPROC2 software developed at the NRO \citep{tsut97} and the AIPS package developed at the National Radio Astronomy Observatory (NRAO). 
The total on-source integration time was 16.4 hours. 
The synthesized beam size was $5''.4 \times 4''.6$ (position angle of $-44^{\degree}$) with natural weighting. 

\section{Results}

\begin{figure}
  \begin{center}
    \FigureFile(73mm,73mm){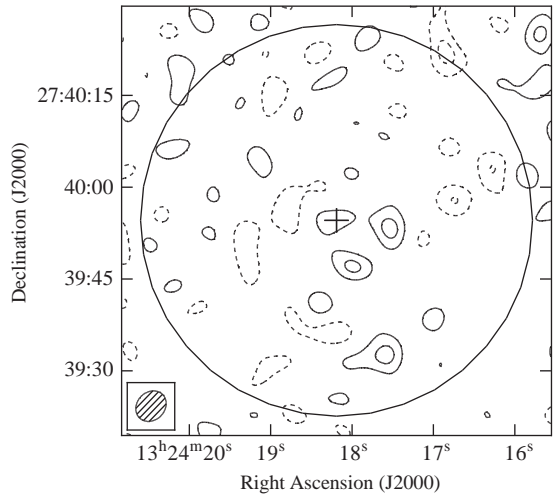} 
  \end{center}
  \caption{
  NMA contour map around the position of SDF-26821 (cross) at 113.63 GHz with a velocity resolution of 84.3 km s$^{-1}$ (1 channel). 
  Contours are $-$2 $\sigma$, $-$1 $\sigma$, 1 $\sigma$, and 2 $\sigma$ (1 $\sigma$ = 5.2 mJy). 
  The large circle represents the field of view ($62''$ diameter). 
  The synthesized beam is shown on the bottom left. 
  }
  \label{fig:map}
\end{figure}

\begin{figure}
  \begin{center}
    \FigureFile(73mm,40mm){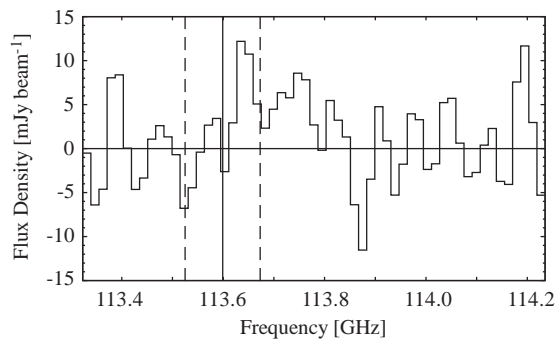} 
  \end{center}
  \caption{
  NMA spectrum at the position of SDF-26821 with a velocity resolution of 84.3 km s$^{-1}$. 
  The vertical solid and dashed lines indicate the frequency range of redshifted CO~(3--2) line expected from the near-infrared spectroscopy. 
  }
  \label{fig:spectrum}
\end{figure}

No significant CO line and continuum emissions are detected with either observations. 
The contour map and spectrum obtained with the NMA are shown in Figures \ref{fig:map} and \ref{fig:spectrum}. 
The rms noise levels for the 45-m and NMA observations are 11 mJy and 6.8 mJy (50 km s$^{-1}$ resolution), respectively. 
The rms noise level of continuum emission with NMA is 1.1 mJy. 
This is more than an order of magnitude higher than expected from the SED model and SFRs, and does not impose tight constraints on the SED. 
The noise level for NMA is measured at the source free region of the contour map avoiding the source signal expected at the phase reference center. 
In what follows, we discuss 2 $\sigma$ limits using the rms noise level obtained with the NMA observations where deeper sensitivity was achieved. 
Following \citet{solo92}, the CO luminosity is
\begin{eqnarray}
L'_{\rm{CO}}= 
3.25\times 10^7 S_{\rm{CO}}\Delta v \nu_{\rm{obs}}^{-2} D_L^2 (1+z)^{-3},
\end{eqnarray}
where $L'_{\rm{CO}}$ is measured in K km s$^{-1}$, $S_{\rm{CO}}$ is the observed CO flux in Jy, $\Delta v$ is a velocity width in km s$^{-1}$, and $D_L$ is the luminosity distance in Mpc. 
Assuming a velocity width of $200\ \rm{km\ s^{-1}}$ based on the CO spectra of BzKs reported by D08, the 2 $\sigma$ upper limit of CO~(3--2) luminosity is $L'_{\rm{CO}} < 3.1 \times 10^{10}$ K km s$^{-1}$ pc$^{-2}$. 

The molecular gas mass is given by $M_{\rm{H_2}} = \alpha_{\rm{CO}} L'_{\rm{CO}}$, where $\alpha_{\rm{CO}}$ is a CO-to-H$_2$ conversion factor. 
We assume that the gas is optically thick and thermalized, and a CO~(3--2)/CO~(1--0) luminosity ratio of unity. 
By adopting a conversion factor of $\alpha_{\rm{CO}}=0.8$ \MO\ (K km pc$^2$)$^{-1}$, the standard value for ULIRGs \citep{down98}, the 2 $\sigma$ upper limit of molecular gas mass is $M_{\rm{H_2}} < 2.5 \times 10^{10}$ \MO. 
This is lower than the median value of $(3.0\pm 1.6) \times 10^{10} \MO$ obtained in the sample of 12 SMGs \citep{grev05}. 

\section{Discussion}
\begin{figure}
  \begin{center}
    \FigureFile(80mm,80mm){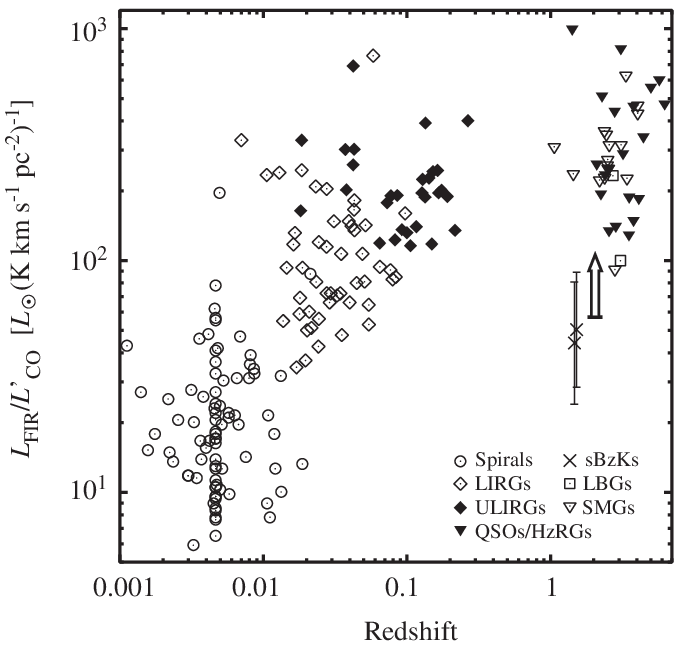} 
  \end{center}
  \caption{
  Star formation efficiency ($L_{\rm{FIR}}/L'_{\rm{CO}}$) as a function of redshift. 
  The 2 $\sigma$ lower limit of SDF-26821 is represented by an upward arrow. 
  For comparison, ratios of various galaxy populations are plotted: 
  local spirals \citep{youn96}, LIRGs, ULIRGs \citep{sand91, solo97}, BzKs (D08), LBGs \citep{bake04, copp07}, 
  SMGs \citep{grev05, dadd08b, tacc08}, QSOs, and HzRGs (\cite{solo05, cari07, maio07, copp08}, and references therein). 
  }
  \label{fig:efficiency}
\end{figure}

The CO luminosity and the FIR luminosity are measures of molecular gas mass and SFR, respectively, and therefore the ratio of $L_{\rm{FIR}}/L'_{\rm{CO}}$ indicates how efficiently stars are formed from molecular gas. 
It is also used as an indicator of star formation efficiency (SFE; \cite{youn86}). 
Figure \ref{fig:efficiency} shows $L_{\rm{FIR}}/L'_{\rm{CO}}$ ratios as a function of redshift using various populations of galaxies;  
local spiral galaxies, 
local luminous infrared galaxies (LIRGs, $L_{\rm{FIR}} = 10^{11-12} \LO$), ULIRGs ($L_{\rm{FIR}} > 10^{12} \LO$), 
sBzKs, Lyman break galaxies (LBGs), SMGs, QSOs, and high-redshift radio galaxies (HzRGs). 
The 2 $\sigma$ lower limit of SDF-26821 is $L_{\rm{FIR}}/L'_{\rm{CO}} > 57$ \LO\ (K km s$^{-1}$ pc$^2$)$^{-1}$.
Considering it is a lower limit, SDF-26821 has a SFE comparable to or higher than those of two sBzKs of D08 ($44^{+37}_{-20}$ and $50^{+39}_{-22}$ \LO\ (K km s$^{-1}$ pc$^2$)$^{-1}$), suggesting that sBzKs could span a wide range of SFEs. 
The ratios of SDF-26821 and D08 sBzKs are higher than the average value for local spirals ($26\pm3$ \LO\ (K km s$^{-1}$ pc$^2$)$^{-1}$), suggesting that sBzKs are not just scaled-up versions of local spirals. 

\begin{figure}
  \begin{center}
    \FigureFile(80mm,80mm){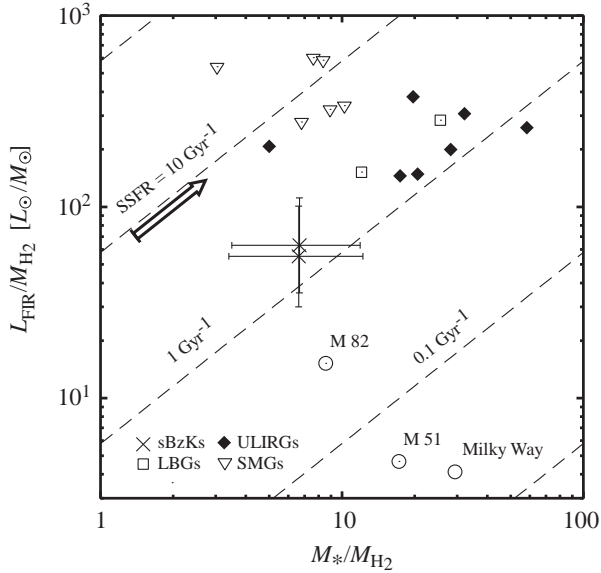} 
  \end{center}
  \caption{
  The ratio of FIR luminosity to molecular gas mass versus the ratio of stellar mass to molecular using samples of distant galaxies (sBzKs, LBGs, and SMGs) and local galaxies (the Milky Way, M~51, M~82, and ULIRGs). 
  The 2 $\sigma$ lower limit of SDF-26821 is represented by an arrow. 
  The $L_{\rm{FIR}}/M_{\rm{H_2}}$ ratio is a measure of star formation efficiency (SFE). 
  Diagonals show constant specific star formation rates (SSFRs; SFRs per unit stellar mass). 
  SFRs are derived from FIR luminosities \citep{kenn98} and stellar masses are converted to a Salpeter IMF. 
  Note that we use dynamical mass for ULIRGs and baryonic mass for M~51 and M~82 instead of stellar mass, and they give upper limits on stellar mass. 
  We use CO-to-H$_2$ conversion factors of $\alpha_{\rm{CO}}=4.6$ \MO\ (K km pc$^2$)$^{-1}$ (Galactic value; \cite{solo91}) for M~51 and M~82, and $\alpha_{\rm{CO}}=0.8$ \MO\ (K km pc$^2$)$^{-1}$ for ULIRGs and distant galaxies.
  The ULIRG samples are taken from \citet{tacc02}, SMG samples are taken from \citet{tacc08} and \citet{dadd08b}, and sBzKs and LBGs are the same as in Figure \ref{fig:efficiency}. 
  The data of FIR luminosities and molecular gas masses are obtained from \citet{chap05} for SMGs of \citet{tacc08}, \citet{solo97} for ULIRGs, \citet{youn96} for M~51 and M~82, and \citet{scov89} and \citet{dame93} for the Milky Way. 
  The stellar masses are taken from \citet{silv98} for M~51 and M~82, and \citet{flyn06} for the Milky Way. 
  }
  \label{fig:stellar_gas-sfe}
\end{figure}

In Figure \ref{fig:stellar_gas-sfe}, we plot the ratio of FIR luminosity to molecular gas mass ($L_{\rm{FIR}}/M_{\rm{H_2}}$, another form of SFE) versus the ratio of stellar mass to molecular gas mass ($M_*/M_{\rm{H_2}}$) of ULIRGs, sBzKs, LBGs, and SMGs. 
For comparison, we show M~51 as a typical spiral, M~82 as a typical starburst, and the Milky Way. 
The 2 $\sigma$ lower limit of SDF-26821 is represented by an arrow. 
In this SFE--$M_*/M_{\rm{H_2}}$ diagram, the ratio of vertical axis to horizontal axis provides specific star formation rates (SSFRs; SFRs per unit stellar mass) by using a conversion from FIR luminosity to SFR \citep{kenn98}. 
The SSFR is thought to be an indicator of current star-forming activity, and the inverse of the SSFR is related to the mass doubling time. 
The SSFR of SDF-26821 (8.8 Gyr$^{-1}$) is a factor of six higher than those of D08 sBzKs (1.4 and 1.6 Gyr$^{-1}$), suggesting that the mass doubling time of SDF-26821 is shorter than those of D08 sBzKs. 
Compared to the other galaxy populations, SDF-26821 and D08 sBzKs have similar SSFRs to the average for the sample of SMGs ($13\pm 4$ Gyr$^{-1}$) and ULIRGs ($2.4\pm 0.8$ Gyr$^{-1}$), respectively, and more than an order of magnitude higher than those of M51 and the Milky Way. 
Taking into account the information of molecular gas mass, the location of a galaxy within this diagram shows the states of star-formation in galaxies (SFE, SSFR, and the fraction of stellar mass to molecular gas mass). 
Figure \ref{fig:stellar_gas-sfe} suggests that SDF-26821 is a different system from D08 sBzKs:
(i) if the actual CO luminosity of SDF-26821 is close to the 2 $\sigma$ upper limit, then 
the SFE is comparable to those of D08 sBzKs. 
On the other hand, the fraction of stellar mass to molecular gas mass of SDF-26821 is a factor of five smaller than those of D08 sBzKs and an order of magnitude smaller than M51 and the Milky Way. 
The high fraction of molecular gas mass to stellar mass in SDF-26821 could relate to an earlier phase of forming stars compared to D08 sBzKs (e.g., \cite{fray97}). 
This is supported by the inferred young age of 50 Myr derived from the SED model fitting (Motohara et al. in prep.). 
(ii) If the actual CO luminosity of SDF-26821 is smaller than the 2 $\sigma$ upper limit, then the SFE is higher than those of D08 sBzKs and closer to SMGs. 
This suggests that SDF-26821 is similar to massively star-forming galaxies at high redshifts and could be a `scaled-down' version (i.e., lower gas and stellar mass) of SMGs . 

Our CO observation reveals that sBzKs could have a wide range of SFEs. 
It is possible that sBzKs are not just scaled-up versions of local spirals and are more like vigorously star-forming galaxies at high redshifts. 
In order to examine the evolution process of these high-redshift star-forming galaxies, more observations of molecular gas are needed. 
However, the sensitivity of current instruments is restricted to only bright sources. 
ALMA will enable us to study the gas environment of normal galaxies at high redshift and to examine the galaxy evolution in terms of molecular gas, which has been less understood thus far.

\bigskip

We would like to acknowledge Takeshi Sakai, Taku Nakajima, and the members of the NRO for observational support. 
We are also grateful to Emanuele Daddi for kindly providing detailed data for the BzKs. 
This work has been supported in part by the Grant-in-Aid for Scientific Research (19540238) from Japan Society for the Promotion of Science (JSPS).
B.~H.\ and M.~H.\ are financially supported by a Research Fellowship from the JSPS for Young Scientists. 


\end{document}